\documentclass{article}

\usepackage[preprint]{neurips_2026}

\usepackage[utf8]{inputenc} 
\usepackage[T1]{fontenc}    
\usepackage[hidelinks]{hyperref}       
\usepackage{url}            
\usepackage{booktabs}       
\usepackage{amsfonts}       
\usepackage{nicefrac}       
\usepackage{microtype}      
\usepackage{xcolor}         

\usepackage{natbib}
\usepackage{graphicx}
\usepackage{float}

\title{Generalized Mean Absolute Directional Loss for Machine Learning Trading Models}

\author{%
  Jakub Michańków \\
  Department of Quantitative Finance and Machine Learning\\
  University of Warsaw\
  \texttt{j.michankow@uw.edu.pl} \\
  \And
   Paweł Sakowski \\
  Department of Quantitative Finance and Machine Learning\\
  University of Warsaw\
  \texttt{p.sakowski@uw.edu.pl} \\
  \And
   Robert Ślepaczuk \\
  Department of Quantitative Finance and Machine Learning\\
  University of Warsaw\
  \texttt{rslepaczuk@wne.uw.edu.pl } \\
}

\begin{document}

\maketitle

\begin{abstract}
    The article presents and evaluates a custom loss function designed specifically for machine learning models used in algorithmic trading. Regardless of the selected asset class and the level of model complexity, the proposed Generalized Mean Absolute Directional Loss (GMADL) function produces superior results and has better numerical properties during optimization than classic regression and classification based loss functions. Better results correspond to higher risk-weighted returns based on buy and sell signals derived from forecasts generated by models trained using GMADL. In practice, GMADL improves model evaluation by aligning the learning objective with trading performance rather than generic error minimization or simple directional accuracy.

    Through additional parameterization, GMADL provides a flexible mechanism for adjusting the loss's sensitivity to different return magnitudes, which affects model evaluation consistency across various market regimes. This improves the selection of model configurations that are more consistent with performance-based criteria.

    Moreover, the implementation uses robust machine learning tools, including frameworks for hyperparameter tuning, architecture testing, and walk-forward optimization, to provide robust and scalable model evaluation across real-world financial data from different asset classes.
\end{abstract}

\section{Introduction}

The main idea of this paper is to improve the methodology of evaluating algorithmic investment strategies, with particular focus on the loss function used during model training and hyperparameter tuning. Within the various identified issues affecting the optimization process, we distinguish the following: selection of adequate training, validation, and testing periods; coverage of diverse asset classes; forward-looking and data-snooping bias in buy/sell signals; lack of sensitivity analysis; and improper loss function choice. The last issue is especially important because it directly shapes the entire optimization process and its outcomes.

Most articles dealing with algorithmic investment strategies (AIS) evaluation focus on specific aspects of testing procedures or empirical results for selected strategies (\cite{wiecki2016all}, \cite{bailey2016probability}, \cite{raudys2016portfolio}, \cite{vo2020high}, \cite{dipersioArtificialNeuralNetworks2016}, \cite{yangDeepLearningStock2019}, \cite{zhang_multi_2018}). We approach this problem more holistically. The central hypothesis verified in this paper is: \textit{The GMADL function more adequately reflects the objective of algorithmic trading strategies by generating better risk-adjusted returns.}

The empirical consequences of this mismatch have been documented directly. The M6 forecasting competition \cite{MAKRIDAKIS2024}, which evaluated both forecast accuracy and investment performance across a large cross-section of participants, found a negligible correlation between the two. Models with high accuracy rankings did not systematically produce better investment outcomes, and vice versa. This finding is consistent with the theoretical argument that point-forecast accuracy and directional profit are structurally different objectives. \cite{DESSAIN2022116970} formalizes this argument, showing that MSE and RMSE are statistically inadequate for evaluating financial ML models and proposing metrics that are explicitly grounded in risk-adjusted returns. \cite{voslepaczuk2021} provides further evidence from a systematic analysis of multiple ML architectures and loss functions, showing that the choice of optimization criterion substantially affects the quality of resulting investment strategies, independently of model architecture.

These findings motivate a work focused specifically on redesigning the loss function for financial ML models. \cite{MichanSakoSlepLSTM} and \cite{MICHANSakoSlepMADL} proposed the Mean Absolute Directional Loss (MADL) as a training objective that rewards correct directional predictions weighted by the magnitude of realized returns, initially in the context of LSTM networks for return forecasting. MADL was subsequently applied to hedging strategy construction (\cite{MichanSakoSlepHedging2024}), showing better resuls compared to traditional loss functions. In these applications, MADL produced better hyperparameter selection and stronger risk-adjusted performance than MSE or MAE. Despite these advantages, MADL relies on sign and absolute value functions, which introduce non-differentiability during optimization. This creates instability in gradient-based optimization — particularly relevant for architectures that depend on smooth gradient flow — and limits the function's adaptability to different return distributions and trading cost structures.

This paper proposes the Generalized Mean Absolute Directional Loss (GMADL), which resolves the numerical limitations of MADL through exponential smoothing and extends its framework with two parameters, $a$ and $b$, that control the slope of the loss function around zero and the degree to which higher realized returns are rewarded during training, respectively. GMADL is fully differentiable, retains the economic alignment of MADL, and provides a flexible mechanism for calibrating the loss function to different asset classes, trading frequencies, and risk profiles. We evaluate GMADL using a transformer model across seven asset classes — equities (SPX, META), currencies (EURUSD), commodities (XAUUSD, ZWF), fixed income (TLT), and cryptocurrency (BTC) — using walk-forward optimization over daily returns spanning 2011 to 2025. Results are assessed using risk-adjusted performance metrics rather than simple forecast accuracy, consistent with the evaluation framework advocated throughout this paper.

\section{Methodology}

\subsection{Mean Absolute Directional Loss Function}\label{MADL. New Loss Function} 

\vspace{5pt}

In \cite{MICHANSakoSlepMADL}, a new loss function was described, the Mean Absolute Directional Loss (MADL), to address common shortcomings in evaluating algorithmic trading strategies.

Authors highlighted that standard error metrics fail to capture the effectiveness of forecasts for generating investment signals. These metrics focus only on forecast accuracy, penalizing all errors equally, regardless of direction, or consider only directional accuracy, ignoring error magnitude. As a result, most studies optimize for metrics that prioritize point forecasts or classification accuracy over investment strategy profitability.

MADL resolves this by better aligning model training to maximize investment strategy profitability. The MADL formula is:

\begin{equation}
\label{MADL}
\textrm{MADL} =\frac{1}{N} \sum_{i=1}^{N} (-1) \times \textrm{sign}({R_{i} \times \hat{R}_{i}}) \times  \textrm{abs}(R_{i} ),
\end{equation}
where $\textrm{MADL}$ is the Mean Absolute Directional Loss, $R_{i}$ is the observed return on interval \(i\), \(\hat{R}_{i}\) is the predicted return on the interval $i$, $\textrm{sign}(X)$ is the function that returns -1,0,1 as the sign of $X$, $\textrm{abs}(X)$ is the function that gives the absolute value of $X$ and $N$ is the number of forecasts.

The function returns the observed return on investment with the predicted direction, allowing the model to evaluate whether the prediction leads to a profit or loss and the actual magnitude of this outcome.

Importantly, the absence of \(\hat{R}_{i}\) in Equation \ref{MADL} is intentional. A properly designed loss function does not require \(\hat{R}_{i}\) because, in algorithmic investment strategies, the critical factor is the prediction's direction ($\textrm{sign}({R_{i} \times \hat{R}_{i}})$) especially on intervals when $\textrm{abs}(R_{i})$ is relatively high.

MADL is specifically designed for generating signals in AIS. When training the model, the function is minimized, ensuring that the strategy yields a profit for negative values and a loss for positive values. MADL is also the primary loss function used for hyperparameter tuning and model estimation.

\begin{figure}[h]
\centering
\includegraphics[width=1\linewidth]{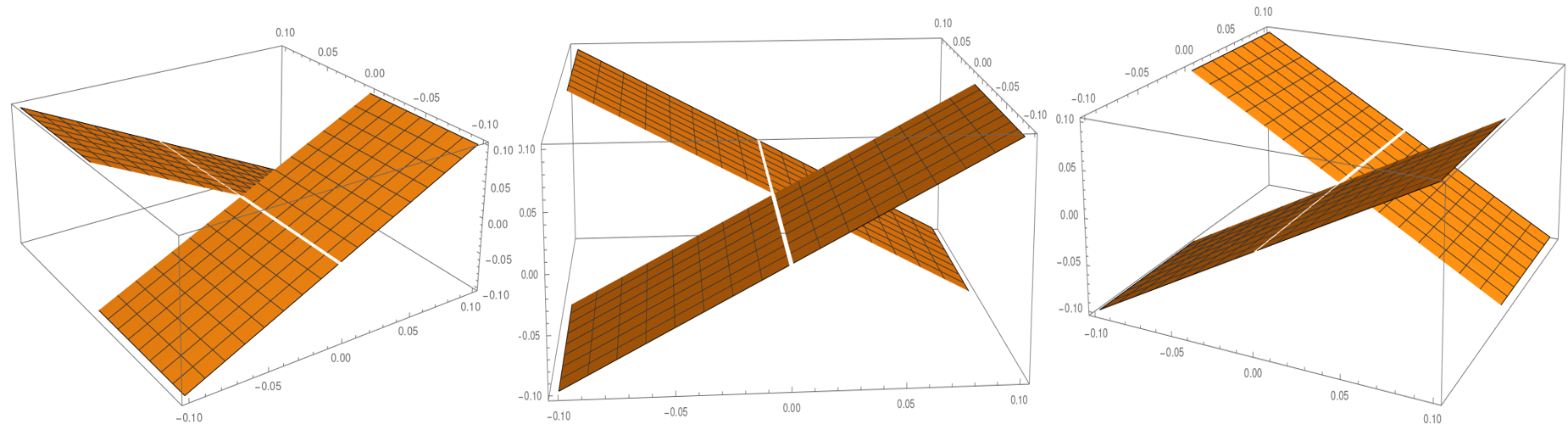}
\footnotesize
Note: x = \(R_i\), y = \(\hat{R}_{i}\), z = MADL
\caption{MADL -- view from different angles in three-dimensional space}
\label{fig:MADLconcept-1}
\end{figure}

Figure \ref{fig:MADLconcept-1} presents the visualization of the MADL function.

Specified like this, MADL provides a tailored loss function optimized for machine learning algorithms in investment strategies. It has proven effective in selecting better hyperparameters for ML models, allowing the development of more profitable and effective investment strategies.

\subsection{Inefficiencies of Mean Absolute Directional Loss Function}\label{InefficiencesofMADL}

\vspace{5pt}

While MADL is effective for optimizing machine learning models in investment strategies, it has notable inefficiencies. One of the primary challenges lies in gradient optimization during the learning process. The reliance on absolute and sign functions introduces non-differentiability at zero, creating complications for smooth gradient-based training. This limitation can affect the convergence of models and affect their overall performance.

One potential direction to address this issue is to explore alternative penalization structures that could result in smoother gradients, making the optimization process more stable and efficient. Such modifications might allow for better alignment of the loss function with the requirements of gradient-based learning algorithms and make the function more reliable and consistent in model training outcomes.

Finally, generalizing the form of MADL could improve its flexibility and applicability. By adapting the loss function to account for the varying actual returns with different rewards, it would become better suited to diverse investment scenarios. This would allow MADL to optimize strategies that rely on frequent trading by including penalization for high trading costs.

Addressing these issues will significantly improve MADL's robustness, making it a more versatile and effective tool for designing profitable investment strategies.

\label{ProposedMethodology}

\subsection{New concept of differentiable and adaptable GMADL function}
\label{new-concept-of-diferrentiable-and-adaptable-gmadl-function}

\vspace{5pt}

Detailed analysis of innovations introduced in MADL and, at the same time, some inconsistencies in its formula allowed us to propose a better loss function: Generalized Mean Absolute Directional Loss (GMADL) function that still focuses on real profit and loss instead of measuring the adequacy of point forecasts however is free from numerical problems that were associated with MADL. The new GMADL formula is as follows:

\begin{equation}
\textsf{GMADL} =\frac{1}{N} \sum_{i=1}^{N} (-1) \times \left( \frac{1}{1+\exp(-a \times {R_{i} \times \hat{R}_{i}})}-0.5 \right) \times \textrm{abs}(R_{i})^{b}
\end{equation} \vspace{-5mm}

\vspace{20pt}

where:

\begin{itemize}
\item
  GMADL is the Generalized Mean Absolute Directional Loss function,
\item
  \(R_{i}\) is the observed return on interval \(i\),
\item
  \(\hat{R}_{i}\) is the predicted return on interval \(i\),
\item
  \(\exp(X)\) is the function which gives the exponential value
  of \(X\),
\item
  \(\textrm{abs}(X)\) is the function which gives the absolute value of
  \(X\)
\item
  \(N\) is the number of forecasts.
\item
  \(a\) - parameter responsible for the slope of GMADL function around 0
\item
  \(b\) - parameter responsible for the rewarding of higher actual
  return in the process of learning
\end{itemize}

The Generalized Mean Absolute Directional Loss (GMADL) function addresses the practical limitations observed in traditional loss functions, particularly those that fail to reflect the profitability goals of algorithmic strategies. By introducing an exponential smoothing term, GMADL ensures differentiability across all points, improving the convergence of gradient-based optimization methods. In addition, the function introduces parameters \textit{a} and \textit{b}, which allow users to tune the balance between directional accuracy and return magnitude, making GMADL adaptable to different trading contexts and frequencies. This novel framework shifts the focus from error minimization to maximizing potential returns, linking theoretical model performance and real-world investment objectives.

\subsection{GMADL function characteristics}
\label{gmadl-function}

\vspace{5pt}

The GMADL function stands out for its ability to capture both the direction of predicted returns and the scale of actual price movements, ensuring that model updates emphasize outcomes with higher economic significance. Unlike conventional metrics that treat all forecast errors equally, GMADL weights errors based on the product of predicted and realized returns, prioritizing instances where the model can capitalize on large price swings. Moreover, by adjusting parameters \textit{a} and \textit{b}, practitioners can amplify or dampen the sensitivity to these swings, making GMADL suitable for various market conditions ranging from stable, low-volatility periods to rapid, high-volatility spikes. This characteristic robustness offers a marked improvement in the alignment of model training with the performance metrics most relevant to algorithmic trading strategies.

In summary, the modified differentiable version of MADL (GMADL) gives us at least several edges:

\begin{itemize}
\item
  differentiability at each point, after introducing the exponential function in the denominator and depending on the value of parameter \textit{a},
\item
  adaptability to investor aims, GMADL introduces the graduate value of reward depending on the distance between \(R_{i}\) and 0
\end{itemize}

which are of high importance for algorithmic investment strategy optimization.

\subsection{GMADL function - differentiability}
\label{gmadl-function---differentiability}

\vspace{5pt}

As mentioned, one of the main strengths of GMADL lies in its differentiability, achieved through an exponential function that smooths out the discontinuities typically associated with sign-based or absolute-value-based loss components. This design choice significantly improves the stability and speed of gradient-based learning algorithms, including those used in sophisticated neural network architectures such as Transformers, LSTMs, or hybrid models. By mitigating the risk of gradient explosions or vanishing updates around zero-return thresholds, GMADL fosters a more reliable training process, resulting in models that converge faster and show greater predictive consistency in both simulated and real-market environments.

From a technical point of view, the following innovations are responsible for the improvements of GMADL compared to MADL:
\begin{itemize}
\item
Introducing the $\exp(X)$ function in the denominator of the GMADL enables differentiating around 0.
\item
  The introduction of the value of \(a\) parameter that determines the slope (the gradient) of the GMADL function around 0.
\end{itemize}

  In order to present the fluctuations of GMADL with a specific set of parameters \textit{a} and \textit{b}, we set \(a\) to a high value (e.g. 1000), because it creates a steeper slope around zero.

\vspace{5pt}

\begin{figure}[h]
\centering
\includegraphics[width=1\linewidth]{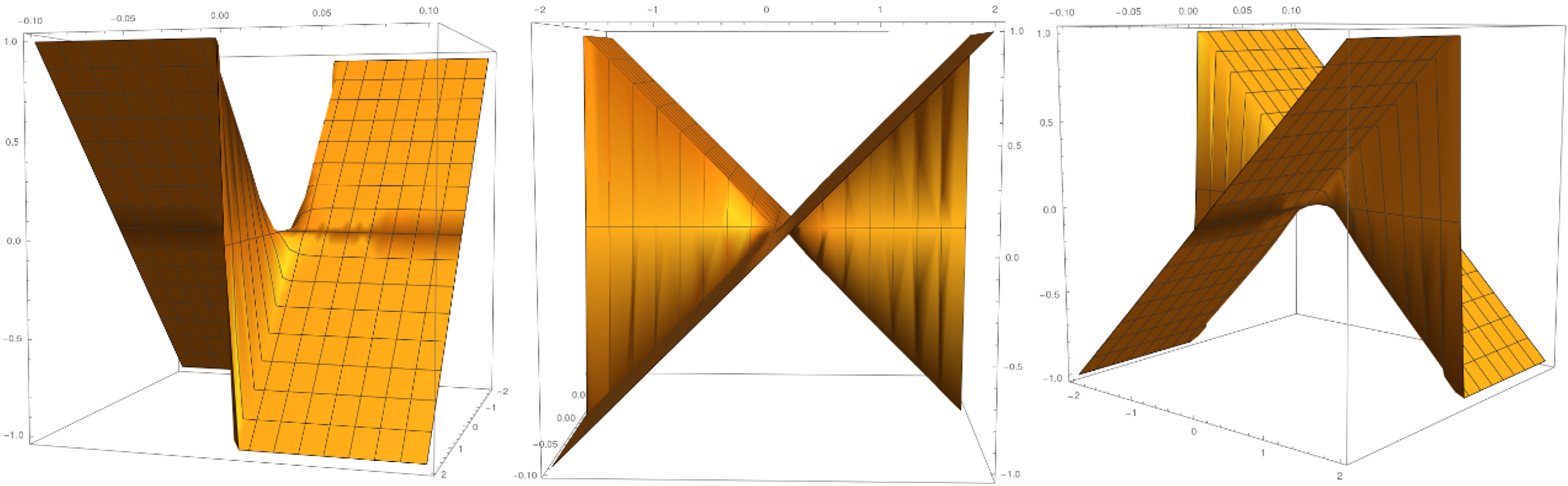}
\footnotesize
Note: x = \(R_i\), y = \(\hat{R}_{i}\), z = GMADL
\caption{a=1000, b=1 -- view from different angles.}
\label{fig:MADLconcept-2}
\end{figure}

\vspace{5pt}

The fluctuations of Figure \ref{fig:MADLconcept-2}  around 0 show how the above-mentioned improvements work.

\subsection{GMADL function - adaptability and scalability}
\label{gmadl-function---adaptability-and-scalability}

\vspace{5pt}

\begin{figure}[h]
\centering
\includegraphics[width=1\linewidth]{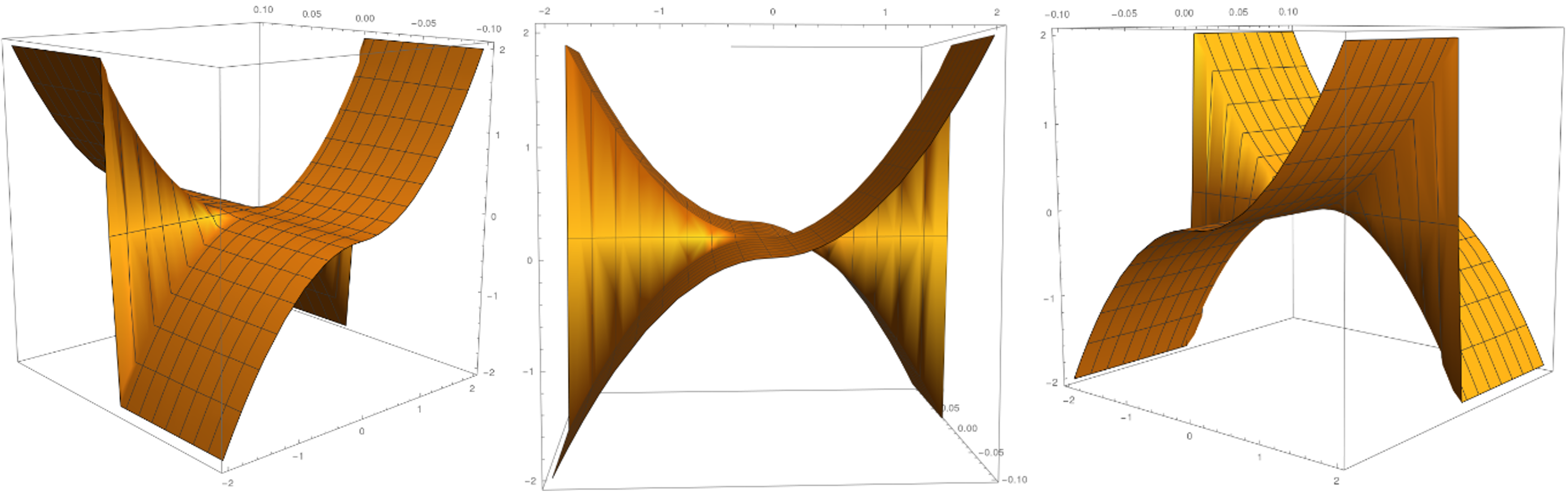}
\footnotesize
Note: x = \(R_i\), y = \(\hat{R}_{i}\), z = GMADL
\caption{a=1000, b=2 -- view from different angles.}
\label{fig:MADLconcept-3}
\end{figure}

\begin{figure}[h]
\centering
\includegraphics[width=1\linewidth]{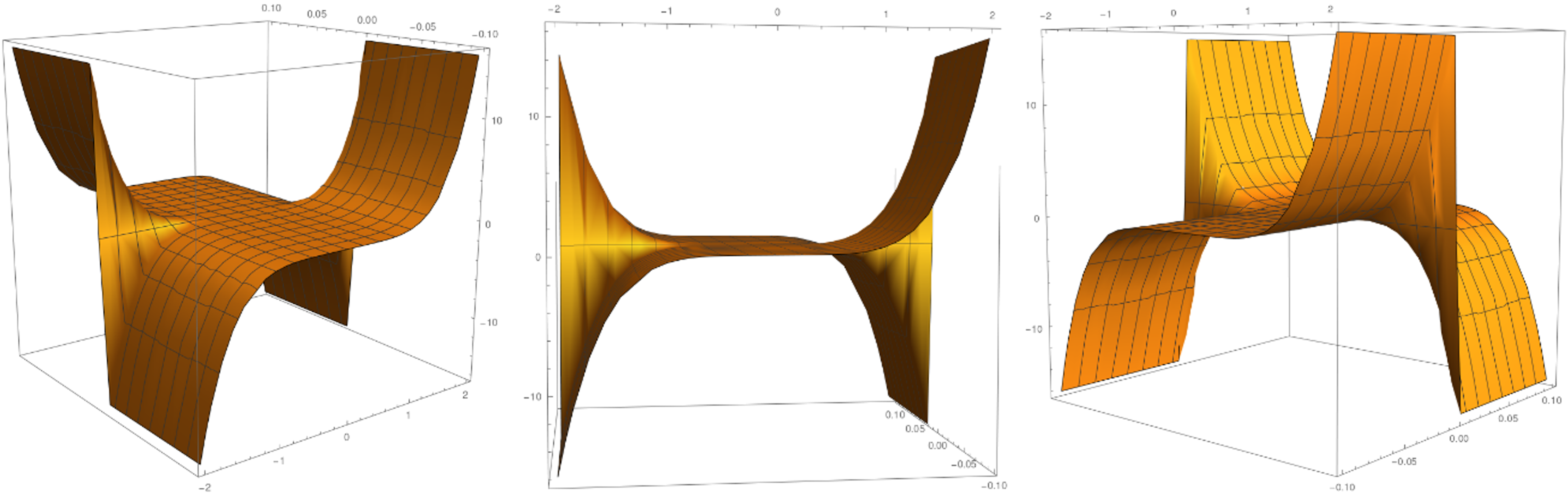}
\footnotesize
Note: x = \(R_i\), y = \(\hat{R}_{i}\), z = GMADL
\caption{a=1000, b=5 -- view from different angles.}
\label{fig:MADLconcept-4}
\end{figure}

Taking into account the transaction costs and the fact that, especially with high frequency data, we incur the substantial value of these transaction costs which significantly affect the performance of the algorithmic investment strategies, we introduced \textbf{parameters \(a\) and \(b\)} which are responsible for additional parameterization of our loss function. If we would like to additionally reward our loss function (minimize it) after the occurrence of higher $R_{i}$ with a higher proportional reward in order to stimulate the process of learning of larger returns we can manipulate with the value of \(b\).   Therefore, the increase of \(b\) parameter results in rewarding of higher actual return in the process of learning. The example of such a change in the shape of the GMADL function can be seen in Figure \ref{fig:MADLconcept-3} and \ref{fig:MADLconcept-4}.

\vspace{5pt}

\vspace{5pt}

\vspace{5pt}

\vspace{5pt}

The adaptability of GMADL parameters responds to the unique demands of different trading horizons, asset classes, and risk appetites. For instance, a larger value of \textit{b} can place greater emphasis on high-return events, which is particularly advantageous in volatile markets where less frequent but large price movements dominate profitability. Similarly, tuning \textit{a} helps calibrate the function’s sensitivity around zero, making it suitable for high-frequency scenarios where small directional errors can accumulate into significant losses due to transaction costs. The result is a highly scalable loss function that.

\subsection{GMADL function - further direction of improvement}
\label{gmadl-function---further-direction-of-improvement}

\vspace{5pt}

Although GMADL offers a robust and flexible framework, several avenues remain for its ongoing refinement. One promising direction involves integrating transaction cost models directly into the loss function, thereby penalizing trades that marginally exceed breakeven thresholds. Additionally, combining GMADL with complementary metrics—such as the Sharpe ratio or drawdown-based measures—could facilitate a more holistic optimization approach that balances growth and downside risk. Future work might also explore dynamically adjusting parameters \textit{a} and \textit{b} in real-time based on market regime detection or machine learning models indications, to enhance the function’s adaptability. By continuing to evolve GMADL in response to emerging market trends and computational techniques, researchers and practitioners can further solidify its role as a cornerstone for efficient, data-driven trading models.

\section{Research description}

In results section we present evaluation of GMADL function using transformer model trained on several different financial asset classes representing broad market dynamics.

\subsection{Data}

We use simple returns of daily frequency for six different financial assets (obtained from public sources at Stooq and Yahoo Finance): Bitcoin, TLT US (20yr Treasury Bonds ETF), ZWF (wheat futures), Gold (XAUUSD), EURUSD, META stock, SNP500,  with data spanning from 03.01.2011 (17.09.2014 for BTC and 18.05.2012 doe META) to 25.03.2025.
\subsection{Model}

The model is a encoder-only transformer consisting of multi-head self-attention followed by a residual connection and batch normalization, then a two-layer MLP (dense-dropout-dense) with another residual connection and batch normalization. The MLP uses Swish activation in the hidden layer and projects back to the input feature dimension for residual compatibility. The final representation is obtained by flattening the dimension, followed by a single linear output neuron without bias. The model is optimized with Adafactor with gradient clipping. These forecasts are then used to create weighted signals used in the trading strategy. 

Input data consist of lagged simple returns up to time \textit{t} while we prediction target is a single \textit{t+1} return. We use walk forwad testing with rolling window, where the train set is 3 years of data (768/1095) and test set is one year (252/365). The entire out of sample consists of approximately ten test windows combined into a single series. Models were trained on Apple Silicon M3 processor, with each full cycle taking around 60 minutes.

\subsubsubsection{Model hyperparameters}

Hyperparameters were tuned using Bayesian optimization with 25 trials per walk-forward window, minimizing validation loss. The sequence length, number of encoder layers, and number of epochs were fixed across all experiments. Table~\ref{tab:hparams} summarizes the search space; the rightmost column reports the most frequently selected values across all assets and windows.
\begin{table}[h]
\centering
\caption{Hyperparameter search space and most frequently selected values.}
\label{tab:hparams}
\begin{tabular}{llll}
\hline
Parameter & Range & Step & Typical best \\
\hline
Head size & 16--128 & 32 & 32 \\
Number of heads & 2--8 & 2 & 2 \\
MLP dimension & 64--512 & 128 & 128 \\
Dropout rate & 0.001--0.3 & 0.1 & 0.05 \\
L2 regularization & {0.00001, 0.0001, 0.001} & -- & 0.0001 \\
Learning rate & 0.001--0.1 & 0.001 & 0.005 \\
\hline
\multicolumn{4}{l}{Fixed: sequence length = 8, encoder layers = 1, epochs = 600, batch size = 32.} \\
\hline
\end{tabular}
\end{table}

\section{Evaluation results}

Table~\ref{tab:all_results} evaluates the transformer model trained with GMADL against a passive Buy\&Hold (B\&H) benchmark across seven assets. Transformer with GMADL outperforms B\&H on five assets — SPX, META, EURUSD, TLTUS, and ZWF — mostly through lower maximum drawdown and positive returns on assets where B\&H yields near-zero or negative annualized returns (EURUSD, TLTUS, ZWF). For XAUUSD, GMADL reduces volatility and drawdown duration but at the cost of lower annualized return and IR. For BTC, B\&H is better on all metrics except maximum drawdown duration, which is consistent with the asset's very strong directional trend over the entire period making it difficult to beat.

\begin{table*}[h]
\centering
\caption{Performance comparison across assets: Buy and Hold (B\&H) vs Transformer with GMADL. Best values are bolded.}
\label{tab:all_results}
\begin{tabular}{llcccccccc}
\toprule
Asset & Strategy & aRC & aSD & MD & MLD & IR* & IR** & IR*** & nObs \\
\midrule

SPX    & B\&H   & 10.86 & \textbf{17.35} & 33.92 & \textbf{2.03} & \textbf{0.63} & 0.20 & 0.01 & 2809 \\
       & GMADL  & \textbf{12.09} & 21.95 & \textbf{24.17} & 2.79 & 0.55 & \textbf{0.28} & 0.01 & 2809 \\
\midrule

META   & B\&H   & \textbf{22.96} & 37.95 & 76.74 & \textbf{2.36} & \textbf{0.60} & 0.18 & 0.02 & 2462 \\
       & GMADL  & 18.68 & \textbf{34.51} & \textbf{49.01} & 2.56 & 0.54 & \textbf{0.21} & 0.02 & 2462 \\
\midrule

EURUSD & B\&H   & -2.07 & \textbf{7.67} & 31.11 & 11.31 & -0.27 & -0.02 & 0 & 2914 \\
       & GMADL  & \textbf{2.68} & 9.70 & \textbf{19.47} & \textbf{4.71} & \textbf{0.28} & \textbf{0.04} & 0 & 2914 \\
\midrule

XAUUSD & B\&H   & \textbf{8.35} & 13.85 & \textbf{23.86} & 5.38 & \textbf{0.60} & \textbf{0.21} & 0 & 2902 \\
       & GMADL  & 4.45 & \textbf{9.72} & 27.97 & \textbf{3.66} & 0.46 & 0.07 & 0 & 2902 \\
\midrule

TLTUS  & B\&H   & -0.09 & 14.92 & 51.12 & 4.62 & -0.01 & 0.00 & 0 & 2808 \\
       & GMADL  & \textbf{6.78} & \textbf{14.53} & \textbf{27.01} & \textbf{3.55} & \textbf{0.47} & \textbf{0.12} & 0 & 2808 \\
\midrule

ZWF    & B\&H   & -0.35 & \textbf{31.50} & 65.06 & 6.95 & -0.01 & 0.00 & 0 & 2808 \\
       & GMADL  & \textbf{9.70} & 41.67 & \textbf{59.46} & \textbf{6.77} & \textbf{0.23} & \textbf{0.04} & 0 & 2808 \\
\midrule

BTC    & B\&H   & \textbf{52.65} & \textbf{69.33} & \textbf{83.40} & 2.96 & \textbf{0.76} & \textbf{0.48} & \textbf{0.09} & 2747 \\
       & GMADL  & 31.88 & 69.54 & 83.68 & \textbf{2.89} & 0.46 & 0.17 & 0.02 & 2747 \\

\bottomrule
\end{tabular}
\end{table*}

\begin{figure}[H]
\centering

\includegraphics[width=\linewidth,height=2.5cm,keepaspectratio]{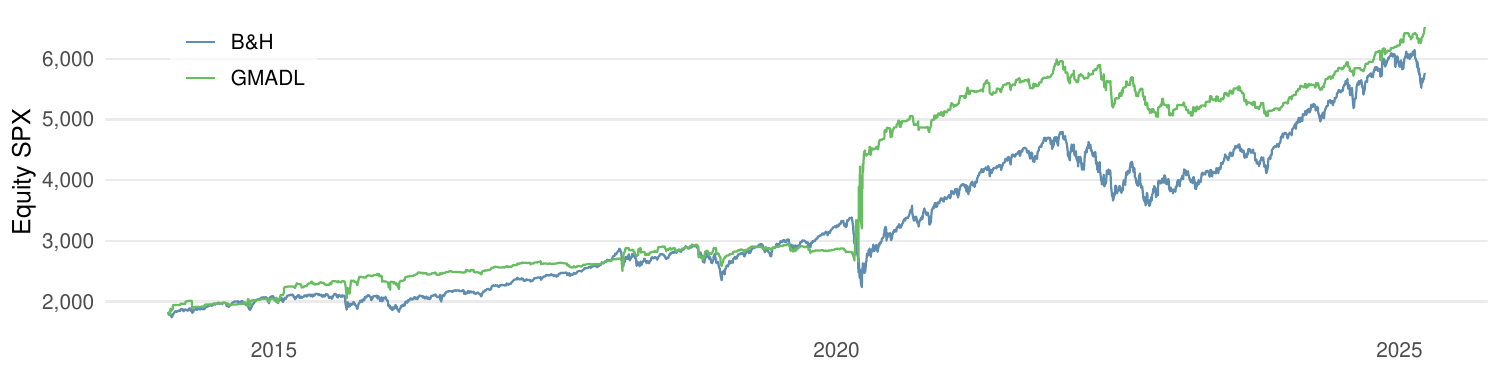}
\vspace{-3mm}

\includegraphics[width=\linewidth,height=2.5cm,keepaspectratio]{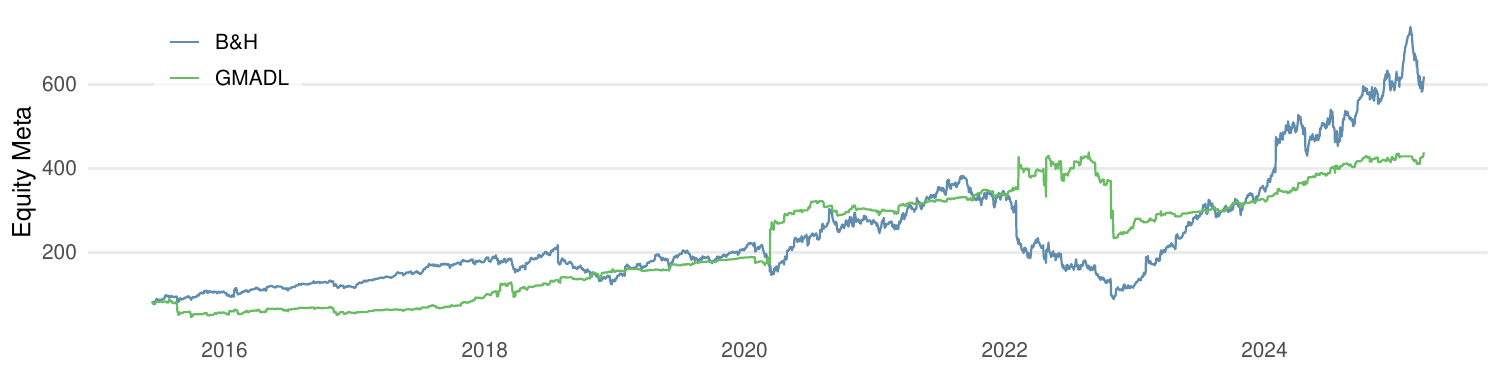}
\vspace{-3mm}

\includegraphics[width=\linewidth,height=2.5cm,keepaspectratio]{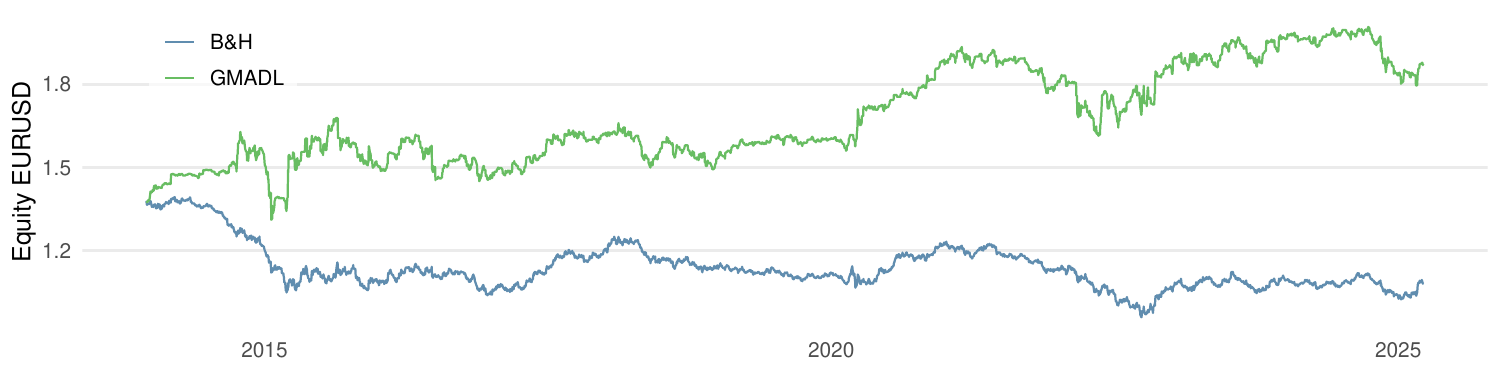}
\vspace{-3mm}

\includegraphics[width=\linewidth,height=2.5cm,keepaspectratio]{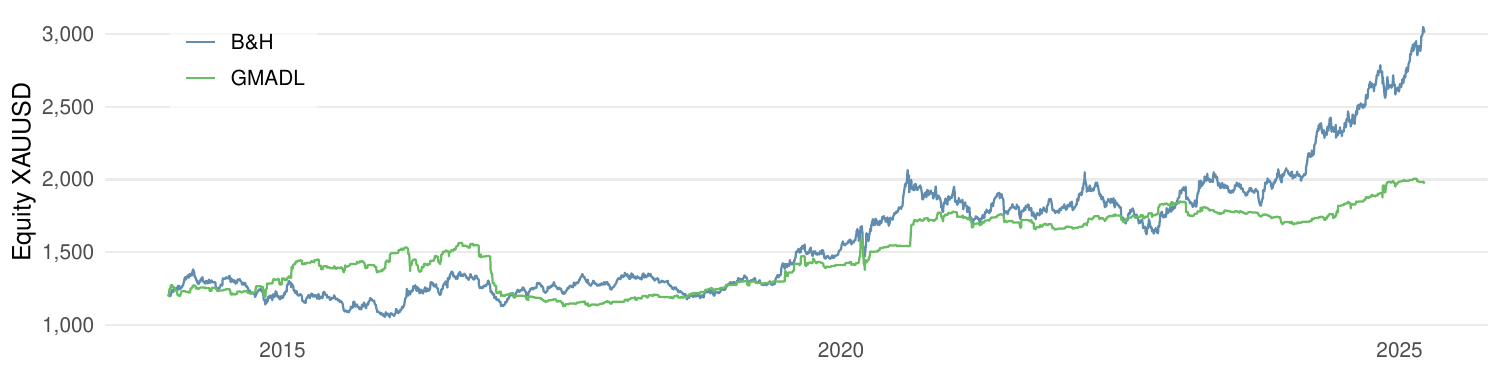}
\vspace{-3mm}

\includegraphics[width=\linewidth,height=2.5cm,keepaspectratio]{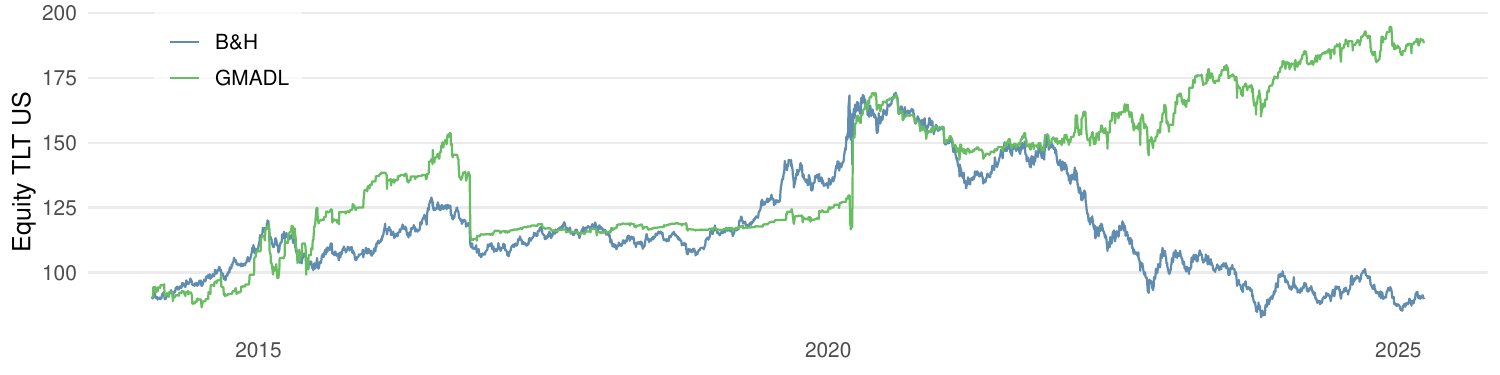}
\vspace{-3mm}

\includegraphics[width=\linewidth,height=2.5cm,keepaspectratio]{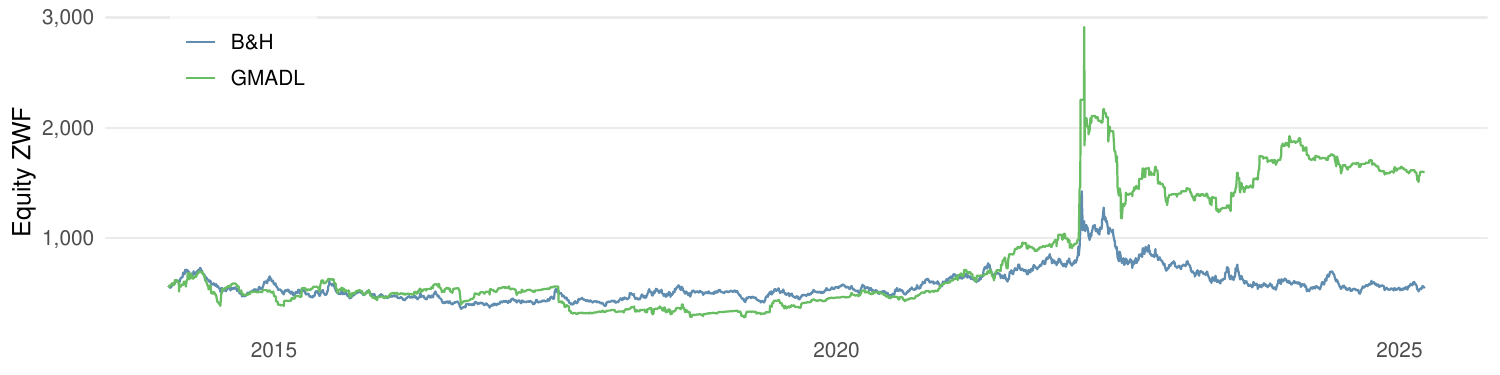}
\vspace{-3mm}

\includegraphics[width=\linewidth,height=2.5cm,keepaspectratio]{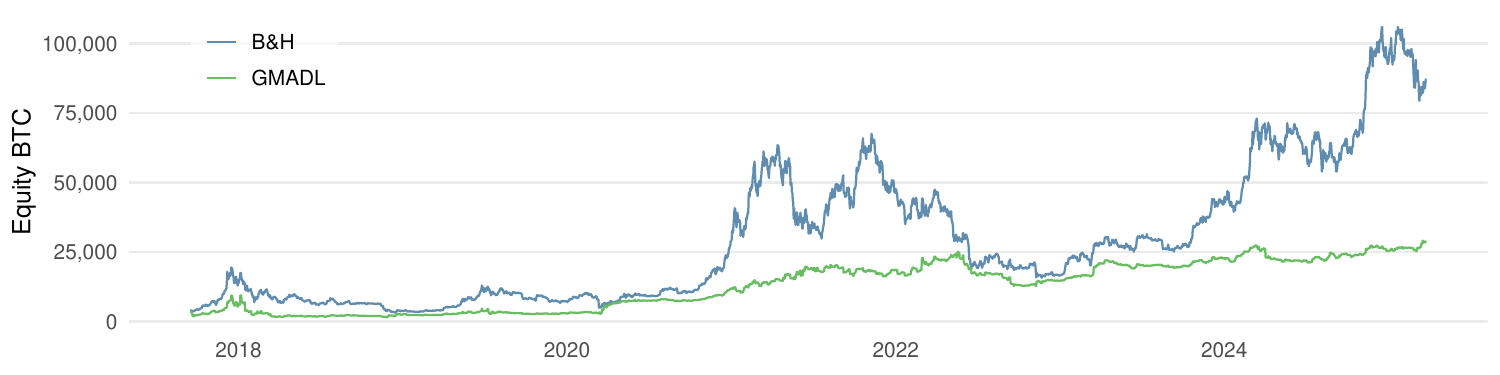}

\caption{Equity curves across assets: B\&H vs GMADL.}
\label{fig:equity_curves}
\end{figure}

Equity curves shown in Figure~\ref{fig:equity_curves} confirm these findings. The GMADL based strategy tracks and exceeds B\&H on SPX throughout the sample, while on EURUSD and TLTUS it generates steady growth against a flat or declining passive position. The ZWF curve shows model capturing the 2022 commodity spike while mostly avoiding the later reversal. On META, both strategies grow comparably until 2022, after which GMADL avoids the sharp drawdown but recovers slowly. The XAUUSD and BTC panels show B\&H pulling ahead in the latter part of the sample which is consistent with the quantitative results.

\section{Conclusions}

The research presented in this paper demonstrates the potential of Generalized Mean Absolute Directional Loss (GMADL) as an effective alternative to conventional loss functions in  algorithmic trading. By addressing the deficiencies of metrics such as MSE, MAE, and even MADL, machine learning models trained with GMADL can better evaluate their objectives with the real-world demands of profitable algorithmic trading strategies. Empirical evaluation across seven asset classes confirms that GMADL-trained strategies outperform a passive benchmark on the majority of tested assets, with consistent improvements in drawdown control and risk-adjusted returns. Using exponential smoothing along with parameters \textit{a} and \textit{b}, GMADL offers a differentiable framework, improving the optimization processes in changing market environments.

These findings underscore the importance of customizing loss functions specifically to investment performance objectives. The ability of GMADL to use directional accuracy in addtition to catching the significant price movements, and include practical constraints such as transaction costs is a step forward in converting theoretical model training to a concrete trading outcomes. It delivers robust gradient behavior, improved adaptability, and scalability for different asset classes and time horizons.

Several limitations should be acknowledged. Walk-forward optimization mitigates but does not eliminate the problem of distributional shift between training and test windows; asset return distributions are non-stationary, and regime changes can cause systematic degradation in out-of-sample performance that no loss function design fully addresses. Additionally, results are reported for a single model architecture; generalizability of GMADL across other architectures remains to be checked.

Future research can build on these findings in several ways. First, integrating more sophisticated transaction cost models directly into GMADL can provide higher control over trading activity, especially in fast-moving markets. Second, combining GMADL with other risk-adjusted performance measures, such as drawdown metrics or the Sharpe ratio, would offer a multidimensional optimization that includes return and risk. Third, extending the evaluation framework to higher frequencies data would provide more empirical evidence for GMADL's generalizability. Lastly, testing the adaptive mechanisms of GMADL—for instance, by dynamically tuning parameters 
\textit{a} and  \textit{b} in response to changing market regimes—could lead to even more effective strategies and give a better understanding of how advanced loss functions can be adjusted to meet the challenges of high-frequency and systematic trading systems.


\newpage
\bibliography{bibfile}{}


\end{document}